\documentstyle[prd,aps,multicol]{revtex}
\begin{document}
\draft
\preprint{IAG/132}
\title{Neutrino sparking and the neutron $\rightarrow$ strange 
stars conversion}

\author{J.E.Horvath$^{1}$ and H.Vucetich$^{2}$}

\address{$^{1}$ Instituto Astron\^omico e Geof\'\i sico, Universidade de
 S\~ao Paulo\\ Av. M.St\'efano 4200, Agua Funda, (04301-904) 
S\~ao Paulo SP Brazil}

\address{$^{2}$ Facultad de Ciencias Astron\'omicas y Geof\'\i sicas, 
Universidad Nacional de La Plata\\ 
Paseo del Bosque S/N , (1900) La Plata, Argentina}
\date{\today}
\maketitle
\begin{abstract}
We address the production of strangelets 
inside neutron stars by means of high-energy neutrino interactions
(sparking). Requiring that neutron stars remain as such along 
their lifetimes, we obtain a bound on the 
probability of a strangelet in the final state and compare it 
with existing laboratory limits. It turns out that this mechanism 
is not likely to drive a neutron $\rightarrow$ strange stars conversion 
for realistic values of the minimum center-mass-energy necessary to 
produce the quark-gluon plasma, a necessary precondition for the 
formation of the strangelet.
\end{abstract}


\begin{multicols}{2}
\def\be{\begin{equation}}
\def\ee{\end{equation}}

\section{Introduction}

A great deal of attention has  been  recently  devoted  to  the 
astrophysical consequences of the strange quark  matter  (SQM) 
hypothesis [1]. Work on different aspects of the birth, 
structure and evolution of the compact strange stars is being done 
with the aim of predicting signatures which may indicate the actual 
existence of this class of objects (see, for example, the reviews 
[2,3,4]). 

As SQM is a low-entropy configuration, the quark gas is  not 
a lower free energy state than an nucleon gas at  intermediate 
temperatures. Compression (i.e. baryochemical  potential $ \mu \not= 0 $) 
is therefore needed to compensate the $-TS$  term  in  the  free 
energy if SQM is to be preferred to NM at $T > 2\, MeV$ [2]. These 
are precisely the physical conditions  generally  believed  to 
exist in young proto-neutron stars [5]  inmediatly  after  the 
passage of the prompt hydrodynamical  shock in  type  II 
supernovae, i.e. inside the  Kelvin-Helmholtz  epoch  of  the 
compact object life. Therefore SQM may play a key role in the very type 
II supernova events as subnuclear energy is released from the conversion 
of neutron to strange matter. This process should result in an 
explosive transient mediated by a detonation front [6,7], a phenomenon 
that would be important for the fate of the collapsed star. Additional 
work [8,9] proved that this conjecture is worth being studied in further 
detail.

A prompt conversion would need the presence of a "dormant" strangelet [10-13] 
or the nucleation of one [10,11,14]. Neither possibility is excluded, 
but also they are not guaranteed. For instance, 
it has been suggested that strong magnetic fields preclude 
SQM nucleation [15], therefore it is interesting to explore other 
alternative mechanisms [10,11]. Generally speaking, these conversion mechanisms 
have been divided in primary (those in which strangelets are produced inside 
the star) and secondary (in which strangelets come from outside).

We shall discuss in the present work the appearance  of  SQM
inside degenerate NM characteristic of neutron stars triggered by neutrino 
sparking. To the best of our knowledge this scenario has not been addressed 
in any detail after the proposal by Alcock {\it et al.} [10] In 
Section II we present the relevant neutrino fluxes and cross-sections
to this problem. A  rough  calculation 
of the strangelet production rate is  given  in 
Section III. Finally, a brief discussion and  conclusions  are 
presented in Section IV.

\section{High-energy neutrinos}

The total cross-sections of neutrinos onto 
nucleons have been recently recomputed by
 Gandhi {\it et al.} [16] using updated parton distributions
in the Altarelli-Parisi framework. They incorporate the latest data
from HERA which goes deeply inside the inelastic regime up to 
$x \, \leq \, 10^{-4}$. Since we are intersted in the reactions of the 
type $\nu N \, \rightarrow \, anything$ we shall employ an "all process"
version of the cross sections $\sigma$ given in [16] and valid above 
a minimum energy $E_{0} \, = \, 1 \, GeV$
\be
10^{-38} {\biggl( {E_{\nu} \over{1 \, GeV}} \biggr)} \, \, cm^{2} \, \, 
\, \, \, \, \, \, \, \; 1 \, GeV < E_{\nu} < 10^{6} \, GeV  
\ee
\be
10^{-36} {\biggl( {E_{\nu} \over{1 \, GeV}} \biggr)}^{0.4}
 cm^{2} \, \, \, \, \, \, \, \; 10^{6} GeV < E_{\nu} < 10^{12} GeV 
\ee

We shall be concerned about inelastic reactions initiated by all neutrino 
flavors. Even though there is considerable uncertainty in the actual 
contribution to the fluxes from various sources, we shall see that our 
final results are largely insensitive to the precise value. 
As in Gandhi {\it et al.} [16] we 
shall adopt the (conservative) differential 
neutrino fluxes $dN_{\nu}/dE_{\nu}$

\be
10^{-6} {\bigl( {E_{\nu} \over{1 \, GeV}} \bigr)}^{-2} GeV^{-1} cm^{2} 
sr s^{-1}   \, \, \, \, \,\,   \, \, \, \, \, E_{\nu} < 10^{6} GeV 
\ee
\be
10 {\bigl( {E_{\nu} \over{1 \, GeV}} \bigr)}^{-3.5} GeV^{-1} cm^{2} 
sr s^{-1}   \, \, \, \, \, \, \, \, \, \, E_{\nu} > 10^{6} GeV . 
\ee

Actually we shall see in the next section that only the first range is relevant 
 for our problem and that a rougher estimate than eqs.(1-4) would have sufficed.

\section{Strangelet production rate}

We wish to address the number of neutrino events that produce a 
strangelet in the final state inside the neutron star. The interesting 
stellar region is that above 
the neutron drip point 
($\rho_{D} \, \simeq \, 4 \, \times \, 10^{11} \, g \, cm^{-3}$); 
since a strangelet could be even produced in the outer shells 
(nuclear lattice) {\it without} necessarily 
triggering the full conversion of the star. Free neutrons do not feel 
the strangelet Coulomb barrier ($\sim 10 - 20 \, MeV$) 
and guarantee the growth and eventually the full burning, justifying the 
focusing on the condition $\rho_{reaction} \, > \, \rho_{D}$. We may say that 
the outer crust acts as a shield against harmful strangelet-producing 
neutrino reactions. If we assume an exponential decrease of the outer 
crust from the drip point up to the surface as a reasonable approximation, 
the optical depth of the outer crust is

\be
\tau (E_{\nu}) \, = \, \sigma (E_{\nu}) \, {\bar{n}} \, 
\delta r \, \; \, ; 
\ee

where ${\bar{n}} \, = \, n_{D} / 4$ and $\delta r \, \sim \, 100 \, m$ is the 
minimal width of this region taken from model calculations.

Let us define the probability of a strangelet production in the final 
state (i.e. the $anything$ of the neutrino interaction) in the 
$\rho \, > \, \rho_{D}$ region as 
$P_{prod} \, = \, P_{QG} \, \times \, P_{s}$, where $P_{QG}$ is the 
probability of a quark-gluon plasma formation and $P_{s}$ is the probability 
of distillating (that is, fragmenting into a reasonable size, separating 
strangeness from antistrangeness and cooling to the ground state [17]; see 
also [18] for a through discussion of the physics of the process) a 
strangelet out of the pre-existing quark-gluon plasma. We have assumed the 
simple parametrized expressions $P_{QG} \, = \, \Theta 
(E_{\nu} -  E^{min})$ and $P_{s} \, = constant$ in our calculation 
(see Ref.[18] and below).
Here $E^{min}$ is the minimum neutrino energy in the laboratory 
frame which would 
yield enough energy density in the center-of-mass to produce 
the quark-gluon plasma 
(which is probably not less than a few $GeV/fm^{-3}$ [19]). 
The strangelet production rate in this approximation is simply

\be
\xi \, = \, 4 \pi R_{NS}^{2} \, \int^{\infty}_{E_{0}} P_{prod} (E_{\nu}) 
{dN_{\nu} \over {dE_{\nu}}} \, \exp (- \tau(E_{\nu})) \, dE_{\nu}
\ee

An approximate integration of eq.(6) yields the result

\be
\xi \, \simeq \, 4 \pi \, 10^{4} \,  P_{s} 
\, \exp (-10 \delta_{100} E^{min}_{GeV}) 
\times {({E^{min}_{GeV}})}^{-2} \, 
\delta_{100}^{-1} \;  ; 
\ee

where we have defined $\delta_{100} \, \equiv \, (\delta r / 100 m)$ 
and $E^{min}_{GeV} \, \equiv \, E^{min}/GeV$. 
If we want that neutron stars remain 
as such along their lives $\tau$, we shall demand the product 
to be $\xi \, {\bar{\tau}} \, \, < \, 1$; or 
(adopting a neutron star mean age of ${\bar{\tau}} \, = \, 1 \, Myr$); that 
$P_{s}$ satisfies the bound

\be
P_{s} \, < \, 6 \, \times \, 10^{-15} \, \delta_{100} \, \times \, 
{({E^{min}_{GeV}} )}^{2} \, 
\exp (10 \delta_{100}{E^{min}_{GeV}}) \, \; 
\ee
 
For reference, an upper bound on the strangelet production rate 
$P_{prod} \, \sim \, 10^{-10}$ has been derived from data 
in heavy ion collisions [19] at ultrarrelativistic center-of-mass 
energies. The main 
uncertainty of our estimate is the actual value $E^{min}$. However, 
for the reactions at energies substantially higher than $1 \, GeV$, the 
strangelet production is exponentially suppressed by the opacity of the 
crust and the conversions can not be triggered by sparking. We note 
in passing that 
this fact justifies the approximation $P_{s} \, = \, constant$ made in 
eqs.(6-8). In other words, that the exact energy dependence of $P_{s}$ 
is irrelevant for our considerations. 

\section{Discussion and conclusions}

We have estimated the strangelet production rate inside neutron 
stars using a simple scheme for understanding that process.
Given its production, the details of a 
quark-gluon fireball production and further 
evolution are quite complicated as discussed in Refs.[17,18]. 
The additional 
complication of the fireball evolution being not in vacuum but in dense 
matter should not really change the situation too much because 
the thermal pressure of the fireball is expected to exceed 
the sum of the vacuum pressure and external pressures. 
However, the main result of this calculation is that neutrino sparking 
is {\it not} likely to be an effective mechanism for the conversions unless 
the minimum neutrino energy for the production of the quark-gluon plasma 
happens to be very low in the stellar environment. Only if that minimum is 
$\sim \, 1 \, GeV$, the astrophysical bound on $P_{s}$ is better than the 
heavy ion one [19] and there is a possibility mantaining 
neutron stars as such without conflicting with laboratory 
limits. For $E^{min}$ as low as 
$ \simeq \, 3 \, GeV$ an effective sparking mechanism would 
require $P_{s} \, \sim \, 1$, which is clearly ruled out. At even higher 
energies neutrinos would eventually reach the required threshold, but in 
this case they will be completely stopped in the outer crust (see eq.8). 
Since no QGP signature is seen in deeply inelastic scattering experiments 
at energies much greater than $E_{\nu} \, \gg 1 \, GeV$, 
we conclude that the process is never effective 
to convert neutron stars into strange stars. 

  It should be kept in mind that the discussed scenario is not 
the only one which can give rise to strangelets in neutron stars.  
Conversion via two-flavor quark matter formation [10] or 
the  presence  of 
strangelets in the supernova progenitor becoming active  after 
neutronization [12,13] are likely alternatives (and there may  be 
another ones as well, see [11]). We conclude that, as long as 
neutrino sparking goes, the case for a mixed NS-SS 
population seems to be weak.
 
\acknowledgements

We wish to acknowledge the PROINTER Program of S\~ao Paulo University 
which made possible our continuing collaboration. The CONICET 
and CNPq Agencies are also acknowledged for partial financial 
support to H.V. and J.E.H. respectively.

\end{multicols}

\begin{references}

\bibitem{Witten}
A.Bodmer, Phys. Rev. D 4, 1601 (1971); 
H. Terazawa INS-Report 336 (1979); S.Chin and 
A.K.Kerman, Phys. Rev. Lett. 43, 1292 (1979);
E.Witten, {Phys. Rev. D} {30}, 272 (1984).

\bibitem{AA}
C.Alcock and A.V.Olinto, {Annu. Rev. Nucl. Part. Sci.} {38}, 161 (1988).

\bibitem{Mac} 
O.G.Benvenuto, H.Vucetich and J.E.Horvath, {Int. Jour. Mod. 
Phys. A}{6}, 4769 (1993). 

\bibitem{Fridolin} 
F.Weber et al., in {\it Proceedings of Vulcano 1996: Frontier Objects 
in Astrophysics and Particle Physics} , eds. F.Giovannelli and G.Mannocchi,
(Ed. Compositore, Bologna, 1997), p. 87. 

\bibitem{SN}
See, for example, A.Burrows and J.Hayes, in {\it Proc. 17th Texas Symp. 
on Relat. Astrophys.}, eds. H.Bohringer, G.E.Morfill  and J.E.Trumper, 
{Ann. N.Y. Academy of Sciences} {759}, 375 (1995).
 
\bibitem{Nos} 
O.G.Benvenuto and J.E.Horvath, {Phys. Rev. Lett.} {63}, 716 (1989) 
; O.G.Benvenuto, J.E.Horvath and H.Vucetich, 
{Int. Jour. Mod. Phys. A} {4}, 257 (1989).

\bibitem{Comb}
O.G.Benvenuto and J.E.Horvath, {Phys. Lett. B}
{213}, 516 (1989).

\bibitem{Gent}
N.Gentile et al., {Astrophys. J.} {410}, 345 (1993); 
S.K.Gosh, S.C.Phatak and P.K.Sahu, {Nuc. Phys. A} {596}, 670 (1995).

\bibitem{Sab}
G.Lugones and O.G.Benvenuto, to appear in Phys. Rev. D (1998).

\bibitem{AFO} 
C.Alcock, E.Farhi and A.V.Olinto, {Astrophys. J.}
{310}, 261 (1986).

\bibitem{Angela}          
A.V.Olinto, in {\it Proc. Strange Quark Matter in Physics and Astrophysics},
eds. P.Haensel and J.Madsen, {Nuc. Phys. B Proc. Supp.} { 24}, 103 (1991).

\bibitem{bomba} 
O.G.Benvenuto and J.E.Horvath, {Mod.Phys.Lett.A} {4}, 1085 (1989).

\bibitem{Glenn} 
N.K.Glenndening, in {\it Proc. Int. Nuc. Phys. Conf.}, eds. M.Hussein 
{et al.} (World Scientific, Singapore 1990), p.711.

\bibitem{Nucl}
J.E.Horvath, O.G.Benvenuto and H.Vucetich, {Phys. Rev. D} {45}, 3865 
(1992); M.L.Olesen and J.Madsen, {Phys. Rev. D} {49}, 2968 (1994); 
H.Heiselberg, in {\it Proc. Int. Symp. Strangeness and Quark Matter}, 
eds. G.Vassiliadis et al. (World Scientific, Singapore 1995), p.338 ; 
K.Iida and K.Sato, astro-ph 9705211.

\bibitem{Somenath}
S.Chakrabarthy, {Phys. Rev. D} {46}, 
1233 (1993).

\bibitem{Ghandi}
R.Gandhi {et al.}, {Astropart. Phys} {5}, 81 (1996).

\bibitem{Greiner} 
C.Greiner, P.Koch and H.Stocker, {Phys. Rev. Lett.} {58}, 1825 (1987).

\bibitem{Hank}
H.Crawford, M.Desai and G.Shaw, in {\it Proc. Strange Quark Matter in
Physics and Astrophysics}, eds. J.Madsen and P.Haensel 
{Nuc. Phys. B  Proc. Supp} {24}, 199 (1991).

\bibitem{Apple}
G.Applequist {et al.}, {Phys. Rev. Lett.} {76}, 3910 (1996).

\end{references}
\end{document}